\documentstyle[12pt,amssymb]{article}
\makeatletter
\advance \textheight by \topskip
\def\eqnarray{\stepcounter{equation}\let\@currentlabel=\theequation
\global\@eqnswtrue
\global\@eqcnt\z@\tabskip\@centering\let\\=\@eqncr
$$\halign to \displaywidth\bgroup\@eqnsel\hskip\@centering
  $\displaystyle\tabskip\z@{##}$&\global\@eqcnt\@ne 
  \hfil$\displaystyle{\hbox{}##\hbox{}}$\hfil
  &\global\@eqcnt\tw@ $\displaystyle\tabskip\z@
  {##}$\hfil\tabskip\@centering&\llap{##}\tabskip\z@\cr}
\@addtoreset{equation}{section}
  \def\theequation{\thesection.\arabic{equation}}
\makeatother
\def\Ad{\mathop{\rm Ad}\nolimits}
\def\p{\partial^{}}
\def\tr{\mathop{\rm tr}\nolimits}
\def\goth{\frak}

\begin{document}

\begin{titlepage}
\hbox to \hsize{\hfil hep-th/9609030}
\hbox to \hsize{\hfil IHEP 96--40}
\hbox to \hsize{\hfil May, 1996}
\vfill
\begin{center}
{\large \bf Hamiltonian reduction of free particle motion} \\
{\large \bf on group SL(2, ${\Bbb R}$)}
\end{center}
\vskip 1cm
\normalsize
\begin{center}
{A. V. Razumov\footnote{E--mail:
razumov@mx.ihep.su} and V. I. Yasnov}\\ 
{\small \it Institute for High Energy Physics, 142284 Protvino, Moscow Region,
Russia}
\end{center}
\vskip 2.cm
\begin{abstract}
\noindent
The structure of the reduced phase space arising in the Hamiltonian
reduction of the phase space corresponding to a free particle motion on
the group ${\rm SL}(2, {\Bbb R})$ is investigated.  The considered
reduction is based on the constraints similar to those used in
the Hamiltonian reduction of the Wess--Zumino--Novikov--Witten model to
Toda systems. It is shown that the reduced phase space is
diffeomorphic either to the union of two two--dimensional planes, or
to the cylinder $S^1 \times {\Bbb R}$.  Canonical coordinates are
constructed for the both cases, and it is shown that in the first
case the reduced phase space is symplectomorphic to the union of two
cotangent bundles $T^*({\Bbb R})$ endowed with the canonical
symplectic structure, while in the second case it is symplectomorphic
to the cotangent bundle $T^*(S^1)$ also endowed with the canonical
symplectic structure.
\end{abstract}
\vfill
\end{titlepage}

\section{Introduction}

Presently there is the well known method of obtaining various Toda
systems by Hamiltonian reduction of the Wess--Zumino--Novikov--Witten
(WZNW) model. In particular, the WZNW model for the case of the Lie
group ${\rm SL}(2, {\Bbb R})$ gives the simplest Toda system, the
Liouville equation.  The main technical tool used in the reduction
under discussion is the Gauss decomposition of the group element
describing the configuration of the WZNW model (see, for example,
\cite{FRRTW92}). Already in an early paper on the Hamiltonian
reduction of the WZNW model \cite{FWBFR89}, it was emphasised that
the Gauss decomposition is local.  Therefore, the reduced phase space
coincides with the phase space of the corresponding Toda model only
locally and may have more complicated global structure.
Unfortunately, one has not yet succeeded to obtain an exhaustive
information on the topology of the reduced phase space for the case
of the total two dimensional WZNW model.

In this connection, in work \cite{TFe95} a one dimensional model
obtained from the WZNW model by the restriction to configuration
which do not depend on the space variable, was considered. Actually, this
is the model describing the motion of a particle on the corresponding
group manifold. The authors of work \cite{TFe95} demonstrated that
for the case of the Lie group ${\rm SL}(2, {\Bbb R})$, in accordance with
the value of the parameters characterising the reduction, there are
two different cases.  In the first case, the reduced phase space
consists of the union of two phase spaces of the Liouville model. In
the second one, the phase space has a more complicated topology. A
detailed, although not exhaustive, investigation of the reduced phase
spaces for various Lie groups was performed in
\cite{FTs95}.

A quantisation of the reduced system for the case of the Lie group
${\rm SL}(2, {\Bbb R})$ was considered in paper \cite{Ful95}.  The
reduced phase space was considered there as a surface glued from two
patches. The quantisation was performed separately for each, then a
procedure of sewing the corresponding wave functions was performed.
Another variant of a quantisation procedure, which is also based on a
local representation of the phase space, is given in
\cite{KTs96}.

{}From our point of view, a quantisation based on the global
description of the phase space would be most convincing. In the
present work we make a first step in such a direction by
demonstrating that the reduced phase space of the model based on the
Lie group ${\rm SL}(2, {\Bbb R})$ is diffeomorphic either to the
union of two planes, or to the cylinder $S^1 \times {\Bbb R}$.  For
both cases we introduce canonical coordinates and show that in the
first case the reduced phase space is symplectomorphic to the union
of two cotangent bundles $T^*({\Bbb R})$ endowed with the canonical
symplectic structure, while in the second case it is symplectomorphic
to the cotangent bundle $T^*(S^1)$ also endowed with the canonical
symplectic structure.  Note that in work \cite{KTs96} it was already
observed that the reduced phase space has the topology described
above, however the differential geometric equivalence was not proved.

\section{Matrix Lie groups}

Let $G$ be a real matrix Lie group, in other words, some Lie
subgroup of the Lie group ${\rm GL}(m, {\Bbb R})$.  Identify the Lie
algebra ${\goth g}$ of the Lie group $G$ with the corresponding
subalgebra of the Lie algebra ${\goth gl}(m, {\Bbb R})$. Denote by
$g = \|g_{ij}\|$ the matrix--valued function on $G$, defined as
\[
g_{ij} (a) = a_{ij},
\]
for any $a = \|a_{ij}\| \in G$.

The Maurer--Cartan form of $G$ is the matrix--valued 1--form, given
by the relation
\begin{equation}
\theta = g^{-1} d g. \label{5}
\end{equation}
The form $\theta$ is invariant with respect to the right shifts in
$G$, takes values in the Lie algebra ${\goth g}$, and satisfies the
equality
\begin{equation}
d \theta + \theta \wedge \theta = 0. \label{1}
\end{equation}

Introduce on $G$ local coordinates $y^\mu$ and represent the
Maurer--Cartan form as
\begin{equation}
\theta = (g^{-1} \partial_\mu g) dy^\mu, \label{10}
\end{equation}
where $\partial_\mu = \partial/\partial y^\mu$. The matrix--valued
functions $g^{-1} \partial_\mu g$ take values in the Lie algebra
${\goth g}$.  Fix some basis $e_\alpha$ of ${\goth g}$. The
corresponding structure constants $c^\gamma_{\alpha \beta}$ of the
Lie algebra ${\goth g}$ are defined by the relations
\[
[e_\alpha, e_\beta] = c^\gamma_{\alpha \beta} e_\gamma.
\]
Expanding $g^{-1} \partial_\mu g$ over the basis $e_\alpha$, we get
\begin{equation}
\theta = e_\alpha \theta^\alpha_\mu dy^\mu. \label{2}
\end{equation}
It is not difficult to get convinced that relation (\ref{1}) is
equivalent to the equalities
\begin{equation}
\p_\mu \theta^\alpha_\nu - \p_\nu \theta^\alpha_\mu +
c^\alpha_{\beta \gamma} \theta^\beta_\mu \theta^\gamma_\nu = 0.
\label{7}
\end{equation}
It can be shown that the matrix $\|\theta^\alpha_\mu \|$ is
nondegenerate.  Denote the matrix elements of the inverse matrix by
$\xi^\mu_\alpha$. Thus, we have
\[
\xi^\mu_\alpha \theta^\beta_\mu = \delta^\beta_\alpha.
\]
Using this relation and equalities (\ref{7}), we obtain
\begin{equation}
\xi^\nu_\alpha \p_\nu \xi^\mu_\beta - \xi^\nu_\beta \p_\nu 
\xi^\mu_\alpha - c^\gamma_{\alpha \beta} \xi^\mu_\gamma = 0.
\label{12} 
\end{equation}

There exists also the left--invariant Maurer--Cartan form $\bar
\theta$, defined by the formula
\begin{equation}
\bar \theta = dg g^{-1}. \label{6}
\end{equation}
The form $\bar \theta$ takes values in the Lie algebra ${\goth g}$
and satisfies the relation
\begin{equation}
d \bar \theta - \bar \theta \wedge \bar \theta = 0. \label{3}
\end{equation}
Using the local coordinates $y^\mu$ and the basis $e_\alpha$, represent
the form $\bar \theta$ as
\begin{equation}
\bar \theta = \partial_\mu g g^{-1} dy^\mu = e_\alpha \bar
\theta^\alpha_\mu dy^\mu. \label{11}
\end{equation}
{}From relation (\ref{3}) it follows that
\begin{equation}
\p_\mu \bar \theta^\alpha_\nu - \p_\nu \bar \theta^\alpha_\mu -
c^\alpha_{\beta \gamma} \bar \theta^\beta_\mu \bar
\theta^\gamma_\nu = 0. \label{4}
\end{equation}
The matrix $\| \bar\theta^\alpha_\mu \|$ is nondegenerate. Therefore,
there exist the matrix $\| \bar \xi^\mu_\alpha \|$ with the functions
$\bar \xi^\mu_\alpha$ determined by the relations
\[
\bar \xi^\mu_\alpha \bar \theta^\beta_\mu = \delta^\beta_\alpha. 
\]
{}From equalities (\ref{4}) we get
\begin{equation}
\bar \xi^\nu_\alpha \p_\nu \bar \xi^\mu_\beta - \bar
\xi^\nu_\beta \p_\nu \bar \xi^\mu_\alpha + c^\gamma_{\alpha
\beta} \bar \xi^\mu_\gamma = 0. \label{13}
\end{equation}

Recall that the adjoint representation for the case of a matrix Lie
group is described by the formula  
\[
\Ad(a) u = a u a^{-1}, \qquad a \in G, \quad u \in {\goth g}.
\]
Comparing definitions (\ref{5}) and (\ref{6}) of the forms $\theta$ and $\bar
\theta$, we obtain
\[
\bar \theta = g \theta g^{-1} = \Ad(g) \circ \theta.
\]
The matrix elements $\Ad^\beta_\alpha(a)$ of the adjoint
representation with respect to the basis $e_\alpha$, are determined by
the equality
\begin{equation}
\Ad(a) e_\alpha = a e_\alpha a^{-1} = e_\beta \Ad^\beta_\alpha(a).
\label{14} 
\end{equation}
Using this equality and relations (\ref{2}) and (\ref{11}), we get
\[
\bar \theta^\alpha_\mu = \Ad^\alpha_\beta (g) \theta^\beta_\mu,
\]
that can be also written as
\begin{equation}
\Ad^\alpha_\beta(g) = \bar \theta^\alpha_\mu \xi^\mu_\beta. \label{19}
\end{equation}

Concluding this section, let us obtain a system of partial
differential equations satisfied by the matrix elements of the
adjoint representation of the group $G$. From equality (\ref{14}) it
follows that
\begin{equation}
\partial_\mu(g e_\beta g^{-1}) = e_\gamma
\partial_\mu(\Ad^\gamma_\beta(g)). \label{15}
\end{equation}
It is easy to get convinced that
\[
\partial_\mu(g e_\beta g^{-1}) = [\partial_\mu g g^{-1}, g e_\beta
g^{-1}]. 
\]
Using the equality
\[
\bar \xi^\mu_\alpha \partial_\mu g g^{-1} = e_\alpha,
\]
which follows from (\ref{11}), we get
\[
\bar \xi^\mu_\alpha \partial_\mu (g e_\beta g^{-1}) = e_\gamma
c^\gamma_{\alpha \delta} \Ad^\delta_\beta(g).
\]
Relation (\ref{15}) now gives
\begin{equation}
\bar \xi^\mu_\alpha \partial_\mu (\Ad^\gamma_\beta(g)) =
c^\gamma_{\alpha \delta} \Ad^\delta_\beta(g). \label{16}
\end{equation}

\section{Free motion on matrix Lie group}

The motion of a point particle on a matrix Lie group $G$ is
described by a mapping sending each instant of time to an element of
$G$. Suppose that the scalar product on ${\goth g}$ defined by the
relation 
\begin{equation}
(u, v) = \tr(u v) \label{8}
\end{equation}
for any $u, v \in {\goth g}$, is nondegenerate. This supposition is
equivalent to the requirement of nondegeneracy of the matrix $c =
\| c_{\alpha \beta} \|$, where 
\begin{equation}
c_{\alpha \beta} = \tr(e_\alpha e_\beta). \label{23}
\end{equation}
Scalar product (\ref{8}) is invariant with respect to the action of
the group $G$ in ${\goth g}$, defined by the adjoint
representation. This invariance leads to the equalities
\begin{equation}
\Ad^\gamma_\alpha(a) \Ad^\delta_\beta(a) c_{\gamma \delta} = c_{\alpha
\beta}. \label{18}
\end{equation}

The free motion on the group $G$ is described by the Lagrangian
\begin{equation}
L = \frac{1}{2} \tr (g^{-1} \dot g g^{-1} \dot g), \label{9}
\end{equation}
where dot means the derivative over $t$.
Lagrangian (\ref{9}) in invariant with respect to the right and left
shifts in the group
$G$.  Using equality (\ref{2}), we get
\[
g^{-1} \dot g = (g^{-1} \partial_\mu g) \dot y^\mu =
e_\alpha \theta^\alpha_\mu \dot y^\mu.
\]
This equality allows to write the expression for the Lagrangian $L$
in terms of the coordinates $y^\mu$ and the velocities $\dot y^\mu$:
\[
L = \frac{1}{2} \dot y^\mu G_{\mu \nu} \dot y^\nu,
\]
where
\begin{equation}
G_{\mu \nu} = \theta^\alpha_\mu c_{\alpha \beta}
\theta^\beta_\nu = \bar \theta^\alpha_\mu c_{\alpha \beta} \bar
\theta^\beta_\nu \label{17}
\end{equation}
are the components of the bi--invariant metric on $G$. The last
equality in (\ref{17}) follows from (\ref{18}).

Consider now the Hamiltonian formulation of the model.  The phase
space in this case is the cotangent bundle $T^*(G)$, endowed with the
canonical symplectic structure. The local coordinates $y^\mu$ on $G$
generate the local canonical coordinates $y^\mu$, $p_\mu$ on
$T^*(G)$, and the canonical symplectic 2--form in terms of these
coordinates has the form
\[
\Omega = d (p_\mu d y^\mu).
\]
This symplectic form leads to the following Poisson brackets for the
coordinate functions $y^\mu$ and $p_\mu$:
\begin{eqnarray*}
&\{y^\mu, y^\nu\} = 0, \qquad \{p_\mu, p_\nu\} = 0,& \\
&\{y^\mu, p_\nu\} = \delta^\mu_\nu. 
\end{eqnarray*}
The Legendre transformation is described in the case under
consideration by the relations
\[
p_\mu = \frac{\partial L}{\partial \dot y^\mu} = G_{\mu \nu} \dot 
y^\mu, 
\]
and we have the following expression for the Hamiltonian of the
system: 
\[
H = \frac{1}{2} p_\mu G^{\mu \nu} p_\nu,
\]
where $G^{\mu \nu}$ are the matrix elements of the matrix inverse to
the matrix $\| G_{\mu \nu} \|$. An explicit expression for $G^{\mu
\nu}$ has the form
\begin{equation}
G^{\mu \nu} =  \xi^\mu_\alpha c^{\alpha \beta} \xi^\nu_\beta =
\bar \xi^\mu_\alpha c^{\alpha \beta} \bar \xi^\nu_\beta, \label{21}
\end{equation}
where $c^{\alpha \beta}$ are the matrix elements of the matrix
inverse to the matrix $\|c_{\alpha \beta}\|$.

Define the functions
\[
j_\alpha = - \xi^\mu_\alpha p_\mu, \qquad \bar \jmath_\alpha = -
\bar \xi^\mu_\alpha p_\mu.
\]
Taking into account (\ref{19}), we obtain
\begin{equation}
j_\alpha = \Ad^\beta_\alpha (g) \bar \jmath_\beta. \label{20}
\end{equation}
{}From (\ref{12}) and (\ref{13}) one gets the following expressions for
the Poisson brackets of the functions $j_\alpha$ and $\bar
\jmath_\alpha$:
\[
\{j_\alpha, j_\beta\} = c^\gamma_{\alpha \beta} j_\gamma, \qquad
\{\bar \jmath_\alpha, \bar \jmath_\beta\} = -c^\gamma_{\alpha \beta}
\bar \jmath_\gamma,
\]
while equations (\ref{16}) and relation (\ref{20}) give
\[
\{j_\alpha, \bar \jmath_\beta\} = 0.
\]
The functions $j_\alpha$ and $\bar \jmath_\alpha$ are the generators
of the left and right shifts in the group $G$. Indeed, from relations
(\ref{10}) and (\ref{2}) we get
\[
\xi^\mu_\alpha \partial_\mu g = g e_\alpha.
\]
This equality allows to show that
\[
\{j_\alpha, g\} = g e_\alpha.
\]
Similarly, using (\ref{11}), we obtain
\[
\{\bar \jmath_\alpha, g\} = e_\alpha g.
\]
As it follows from (\ref{21}), the Hamiltonian of the system in terms
of the functions $j_\alpha$ or $\bar \jmath_\alpha$ has the following
form
\begin{equation}
H = \frac{1}{2} j_\alpha c^{\alpha \beta} j_\alpha = \frac{1}{2} \bar
\jmath_\alpha c^{\alpha \beta} \bar \jmath_\beta. \label{33}
\end{equation}

Using (\ref{2}), we get
\[
dy^\mu = \xi^\mu_\alpha c^{\alpha \beta} \tr(e_\beta \theta).
\]
Hence, there is valid the equality
\begin{equation}
\Omega = - \tr(j \theta), \label{42}
\end{equation}
where we have introduced the notation
\[
j = j_\alpha c^{\alpha \beta} e_\beta.
\]
Similarly, we obtain
\[
\Omega = -\tr(\bar \jmath \bar \theta),
\]
where
\[
\bar \jmath = \bar \jmath_\alpha c^{\alpha \beta} e_\beta.
\]
{}From formulas (\ref{20}) and (\ref{18}) it follows that the matrix
valued functions $j$ and $\bar \jmath$ are related by the equality
\begin{equation}
\bar \jmath = g j g^{-1} = \Ad(g) \circ j. \label{22}
\end{equation}

\section{Lie group SL(2, ${\Bbb R}$)}

The Lie group ${\rm SL}(2, {\Bbb R})$ is formed by all real 
$2 \times 2$ matrices 
\[
a = \left(\begin{array}{cc}
a_{11} & a_{12} \\
a_{21} & a_{22}
\end{array}\right), 
\]
satisfying the relation
\[
\det a = a_{11} a_{22} - a_{12} a_{21} = 1.
\]
The Lie algebra of the group ${\rm SL}(2, {\Bbb R})$ is the Lie
algebra ${\goth sl}(2, {\Bbb R})$, which consists of all real traceless
$2 \times 2$ matrices. Consider the canonical basis of ${\goth sl}(2,
{\Bbb R})$: 
\[
e_1 = x_- = \left( \begin{array}{cc}
0 & 0 \\
1 & 0
\end{array} \right), \qquad
e_2 = h = \left( \begin{array}{cc}
1 & 0 \\
0 & -1
\end{array} \right), \qquad
e_3 = x_+ = \left( \begin{array}{cc}
0 & 1 \\
0 & 0
\end{array} \right).
\]
With such a choice of a basis, in accordance with (\ref{23}) we have
\begin{equation}
\| c_{\alpha \beta} \| = \left( \begin{array}{ccc}
0 & 0 & 1 \\
0 & 2 & 0 \\
1 & 0 & 0
\end{array} \right), \qquad 
\| c^{\alpha \beta} \| = \left( \begin{array}{ccc}
0 & 0 & 1 \\
0 & 1/2 & 0 \\
1 & 0 & 0
\end{array} \right). \label{34}
\end{equation}
Introducing the notation
\begin{eqnarray*}
&j_- = j_1, \qquad j_0 = j_2, \qquad j_+ = j_3, & \\
&\bar \jmath_- = \bar \jmath_1, \qquad \bar \jmath_0 = \bar
\jmath_2, \qquad \bar \jmath_+ = \bar \jmath_3,
\end{eqnarray*}
we get for the matrix valued functions $j$ and $\bar \jmath$ the
expressions 
\begin{equation}
j = \left( \begin{array}{cc}
j_0/2 & j_- \\
j_+ & -j_0/2
\end{array} \right), \qquad
\bar \jmath = \left( \begin{array}{cc}
\bar \jmath_0/2 & \bar \jmath_- \\
\bar \jmath_+ & -\bar \jmath_0/2
\end{array} \right). \label{43}
\end{equation}
Using now relation (\ref{22}), we obtain
\begin{eqnarray*}
&\bar \jmath_- = (g_{11})^2 j_- - g_{11} g_{12} j_0 - (g_{12})^2 j_+,& \\
&\bar \jmath_0 = - 2 g_{11} g_{21} j_- + (g_{11} g_{22} + g_{12}
g_{21}) j_0 + 2 g_{22} g_{12} j_+,& \\
&\bar \jmath_+ = - (g_{21})^2 j_- + g_{22} g_{21} j_0 + (g_{22})^2 j_+.&
\end{eqnarray*}
For the case under consideration we have the following expression for
the inverse element:
\[
a^{-1} = \left(\begin{array}{cc}
a_{22} & -a_{12} \\
-a_{21} & a_{11}
\end{array} \right).
\]
{}From this equality we get the explicit expression for the
Maurer--Cartan form:
\[
\theta = \left( \begin{array}{cc}
g_{22} dg_{11} - g_{12} dg_{21} & g_{22} dg_{12} - g_{12} dg_{22} \\
-g_{12} dg_{11} + g_{11} dg_{12} & - g_{21} dg_{12} + g_{11} dg_{22}
\end{array} \right).
\]
Now, taking into account (\ref{42}) and (\ref{43}), we obtain
\begin{eqnarray}
\Omega = && d[j_- (g_{21} dg_{11} - g_{11} dg_{21}) + j_0 (g_{11}
dg_{22} - g_{22} dg_{11} \nonumber \\
&& {} + g_{12} dg_{21} - g_{21} dg_{12})/2 + j_+ (g_{12} dg_{22} -
g_{22} dg_{12})].  \label{31}
\end{eqnarray}

\section{Reduced phase space}

The phase space reduction in question is performed by imposing two
first class constraints: 
\begin{equation}
j_+ = \mu, \qquad \bar \jmath_- = \nu, \label{44}
\end{equation}
where $\mu$ and $\nu$ are two nonzero real constants.  Constraints
(\ref{44}) generate on the surface defined by them an action of a two
dimensional Lie group. This action is free and gives the foliation of
the constraint surface into two dimensional orbits.  As it follows
from the general theory of the Hamiltonian reduction, the space of
orbits is a symplectic manifold with the symplectic structure
inherited from the initial phase space. This phase space is usually
called the reduced phase space.  In the case under consideration the
reduced phase space can be realised as the surface obtained as the
intersection of the constraint surface with the surface defined by
the relations called gauge conditions. Note that the projection of
the considered group action to the configuration space is not a free
action, therefore, the gauge conditions must be imposed on the
coordinates $j_\alpha$ and $\bar \jmath_\alpha$.  It is not difficult
to get convinced that as admissible gauge conditions we can take, for
example, the relations
\[
j_0 = 0, \qquad \bar \jmath_0 = 0.
\]
Taking these gauge conditions into account, we see that the reduced
phase space is defined by the equations
\begin{eqnarray}
&(g_{11})^2 j_- - (g_{12})^2 \mu = \nu, \label{24} \\
&g_{22} g_{12} \mu - g_{11} g_{21} j_- = 0, \label{25} \\
&g_{11} g_{22} - g_{12} g_{21} = 1. \label{26} 
\end{eqnarray}
Multiplying (\ref{25}) by $g_{11}$ and using (\ref{24}) and (\ref{26}),
we get
\begin{equation}
g_{12} \mu - g_{21} \nu = 0. \label{27}
\end{equation}
Substituting this equality into (\ref{25}), we come to the equality
\begin{equation}
g_{12}(g_{22} \nu - g_{11} j_-) = 0. \label{28}
\end{equation}
{}From the other hand, taking into account (\ref{27}), we can rewrite
(\ref{24}) in the form
\[
(g_{11})^2 - g_{12} g_{21} \nu = \nu,
\]
that gives after the usage of (\ref{26}) the relation
\begin{equation}
g_{11}(g_{22} \nu - g_{11} j_-) = 0. \label{29}
\end{equation}
Since the functions $g_{12}$ and $g_{11}$ cannot take zero value
simultaneously, then it follows from (\ref{28}) and (\ref{29}) that
\begin{equation}
g_{22} \nu - g_{11} j_- = 0. \label{30}
\end{equation}
Using relations (\ref{27}) and (\ref{30}) to exclude the functions
$g_{21}$ and $g_{22}$, it is easy to get convinced that system of
equations (\ref{24})--(\ref{26}) is equivalent to the single equation
(\ref{24}).  In other words, the reduced phase space can be
considered as a two dimensional surface in the three dimensional
space with the coordinates $g_{11}$, $g_{12}$ and $j_-$, defined by
equation (\ref{24}). As it follows from (\ref{31}) the symplectic
form on the reduced phase space is given by the relation
\begin{equation}
\Omega = \frac{\mu}{\nu} d[ 2j_- (g_{12} dg_{11} - g_{11} dg_{12}) +
g_{11} g_{12} dj_-]. \label{35}
\end{equation}
The reduced phase space has different topologies depending on the
value of the parameters $\mu$ and $\nu$. Actually, there are two
essentially different variants, determined by the relative sign of
these parameters.

Suppose first that the parameters $\mu$ and $\nu$ have the same sign.
Without loss of generality one can put $\mu = \nu = 1$.  From
(\ref{24}) it follows that in the case under consideration $g_{11}$
cannot take zero value, and the coordinate $j_-$ can be expressed
through $g_{11}$ and $g_{12}$:
\begin{equation}
j_- = \frac{(g_{12})^2 + 1}{(g_{11})^2}. \label{32}
\end{equation}
Thus, the reduced phase space is topologically the union of two
nonintersecting two dimensional surfaces.  It is clear also that
these surfaces can be realised as the open subsets of the plane
described by the coordinates $g_{11}$ and $g_{12}$, singled out by
the conditions $g_{11} > 0$ and $g_{11} < 0$.

{}From (\ref{35}), using (\ref{32}), we get for the symplectic form on
the reduced phase space the following expression
\[
\Omega = - 2 \frac{dg_{12} \wedge dg_{11}}{(g_{11})^2}.
\]
Now it is not difficult to introduce canonical coordinates assuming,
for example, 
\[
Q = \ln g_{11}, \qquad P = -2 \frac{g_{12}}{g_{11}}.
\]
for the case $g_{11} > 0$.  From (\ref{33}) and (\ref{34}) we
conclude that the Hamiltonian of the reduced system coincides with
$j_-$. Taking into account (\ref{32}) and the equalities
\[
g_{11} = \exp Q, \qquad g_{12} = - \frac{1}{2} P \exp Q,
\]
we get
\[
H = \frac{1}{4} P^2 + \exp(-2Q).
\]
The solution of the equations of motions for the system with this
Hamiltonian is well known, therefore we will not write it here.

Consider now the case when the parameters $\mu$ and $\nu$ have
different signs and put $\mu = -\nu = -1$. Introduce for $g_{11}$ and
$g_{12}$ the polar coordinates,
\[
g_{11} = R \sin \Phi, \qquad g_{12} = R \cos \Phi.
\]
{}From equation (\ref{24}) it follows that
\[
R^2 = \frac{1}{j_- \sin^2 \Phi + \cos^2 \Phi}.
\]
We can treat $\Phi$ and $j_-$ as coordinates on the reduced phase
space.  It is easy to see that these coordinated take only those
values for which
\[
j_- \sin^2 \Phi + \cos^2 \Phi > 0. 
\]
The symplectic form in terms of the coordinates 
$\Phi$ and $j_-$ has the form 
\[
\Omega = - \frac{1}{j_- \sin^2 \Phi + \cos^2 \Phi} d j_- \wedge d
\Phi. 
\]
Thus, we see that the coordinates $\Phi$ and $j_-$ have two essential
deficiencies. First, these coordinates do not take arbitrary values,
and, second, they are not canonical.  It is easy to get convinced
that the general form of a coordinate conjugated to $\Phi$ is given
by the relation
\[
\Pi = - \frac{1}{\sin^2 \Phi} \ln \left( j_- \sin^2 \Phi +
\cos^2 \Phi \right) + F(\Phi),
\]
where $F(\Phi)$ is an arbitrary periodic function. Choosing $F(\Phi)
= 0$, we come to convenient canonical coordinates $\Phi$ and $\Pi$.
Note that these coordinates take arbitrary values.

It is not difficult to show now that $g_{11}$, $g_{12}$ and $j_-$ are
expressed via the coordinates $\Phi$ and $\Pi$ in the following way:
\begin{eqnarray}
&g_{11} = \exp(\Pi \sin^2 \Phi/2) \sin \Phi,& \label{36} \\
&g_{12} = \exp(\Pi \sin^2 \Phi/2) \cos \Phi,& \label{37} \\
&j_- = \left( \exp(- \Pi \sin^2 \Phi) - \cos^2 \Phi \right)/\sin^2
\Phi.& \label{38}  
\end{eqnarray}
Relations (\ref{36})--(\ref{38}) give a parametric representation of
the reduced phase space considered as the surface defined by equation
(\ref{24}). Here the function entering this representation are
infinitely differentiable. Moreover, the corresponding tangent
vectors are linearly independent. Therefore, the reduced phase space
is diffeomorphic to the cylinder $S^1 \times {\Bbb R}$.  Taking into
account the fact that the coordinates $\Phi$ and $\Pi$ are canonical
coordinates, we see that the reduced phase space is symplectomorphic
to the cotangent bundle $T^*(S^1)$ endowed with the canonical
symplectic structure.

The Hamiltonian of the system in terms of the coordinates 
$\Phi$ and $\Pi$ is
\[
H = \frac{1}{\sin^2 \Phi} \left( \cos^2 \Phi - \exp(- \Pi \sin^2
\Phi) \right).
\]
Hence, the Hamiltonian equations of motion have the form
\begin{eqnarray}
&\dot \Phi = \exp(- \Pi \sin^2 \Phi),& \label{39} \\
&\dot \Pi = 2 \cot \Phi \left( \frac{1}{\sin^2 \Phi} - \exp(- \Pi
\sin^2 \Phi) \left( \Pi + \frac{1}{\sin^2 \Phi} \right) \right).&
\label{40} 
\end{eqnarray}
Despite of the fact that these equations have a rather complicated
form, they can be easily solved. Indeed, the Hamiltonian of the
system is a conserved quantity, which has the sense of energy.  Let
us look for the solutions of the equations of motion for which $H =
\epsilon$. Using (\ref{38}) and (\ref{39}), we come to the relation
\[
\epsilon = \frac{1}{\sin^2 \Phi} (\cos^2 \Phi - \dot \Phi).
\] 
Consider the function $T = \tan \Phi$. For this function we get the
equation 
\begin{equation}
\dot T = 1 - \epsilon T^2. \label{41}
\end{equation}
In the case $\epsilon < 0$, this equation has the solution
\def\T{\tan \left (\sqrt{- \epsilon} (t - c) \right)}
\def\TS{\tan^2 \left (\sqrt{- \epsilon} (t - c) \right)}
\[
T(t) = \frac{1}{\sqrt{-\epsilon}} \T,
\] 
where $c$ is the integration constant. From this we easily get
\begin{eqnarray*}
&&\cos 2 \Phi(t) = \frac{\epsilon + \TS}{\epsilon - \TS}, \\ 
&&\sin 2 \Phi(t) = - \frac{2 \sqrt{-\epsilon} \T}{\epsilon - \TS}, \\
&&\Pi(t) = \frac{\TS - \epsilon}{\TS} \ln \left[ \frac{\epsilon
+ \epsilon \TS}{\epsilon - \TS} \right].
\end{eqnarray*}
In the case $\epsilon > 0$ equation (\ref{41}) has the solution
\def\TH{\tanh \left (\sqrt{\epsilon} (t - c) \right)}
\def\THS{\tanh^2 \left (\sqrt{\epsilon} (t - c) \right)}
\[
T(t) = \frac{1}{\sqrt{\epsilon}} \TH,
\]
and we come to the relations
\begin{eqnarray*}
&&\cos 2 \Phi(t) = \frac{\epsilon - \THS}{\epsilon + \THS}, \\ 
&&\sin 2 \Phi(t) = \frac{2 \sqrt{\epsilon} \TH}{\epsilon + \THS}, \\ 
&&\Pi(t) = - \frac{\THS + \epsilon}{\THS} \ln \left[ \frac{\epsilon
- \epsilon \THS}{\epsilon + \THS} \right].
\end{eqnarray*}
Thus, we have the explicit solution of the equations of motions.

To construct the corresponding quantum theory, we have to solve the
ordering problem for the quantum Hamiltonian. In any case, in the
representation where the state space is realised as the space of
square integrable functions on the circle, the corresponding operator
is not local. However, the availability of the explicit general
solution to the classical equations of motions gives us a hope that
the eigenvalue problem for the quantum Hamiltonian can be solved.

\section{Conclusion}

In the present work we have studied in detail the structure of the
reduced phase space arising as the result of the Hamiltonian
reduction of the phase space corresponding to a free particle motion
on the Lie group ${\rm SL}(2, {\Bbb C})$. We have shown that the
reduced phase space is diffeomorphic either to the union of two
cotangent bundles $T^*({\Bbb R})$, or to the cotangent bundle
$T^*(S^1)$. Moreover, for both cases we have constructed canonical
coordinates and have shown that the arising phase spaces are 
symplectomorphic to the corresponding cotangent bundles endowed with
the canonical symplectic structure.

Note that despite of a complicated structure of the Hamiltonian,
arising in the second case, the classical equations of motions can be
explicitly integrated. This fact allows us to hope that the quantisation
problem has an explicit solution.

The authors are deeply grateful to G.~L.~Rcheulishvili, G.~P.~Pron'ko and
M.~V.~Saveliev for the interesting and fruitful discussions. This work
was partially supported by the Russian Foundation for Basic Research
(project \# 95--01--00125a).


\small

\begin{thebibliography}{**}

\bibitem{FRRTW92}
L. Feh\'er, L. O'Raifeartaigh, P. Ruelle, I. Tsutsui and A. Wipf,
{\it Phys.\ Rep.} {\bf 222}, 1 (1992).

\bibitem{FWBFR89}
P. Forg\'acs, A. Wipf, J. Balog, L. Feh\'er and L. O'Raifeartaigh,
{\it Phys.\ Lett.} {\bf 227B}, 214 (1989).

\bibitem{TFe95}
I.\ Tsutsui and L.\ Feh\'er,
{\it Prog.\ Theor.\ Phys.\ Suppl.} {\bf 118}, 173 (1995).

\bibitem{Ful95}
T.\ F\"ul\"op,
{\it J. Math. Phys.} {\bf 37}, 1617 (1996).

\bibitem{FTs95}
L.\ Feh\'er and I.\ Tsutsui,
{\it  Regularization of  Toda lattices by Hamiltonian reduction},
 preprint INS-1123 (hep-th/9511118).

\bibitem{KTs96}
H. Kobayashi and I. Tsutsui, {\it Quantum mechanical Liouville model
with attractive potential}, preprint INS-1124 (hep-th/9601111).

\end{thebibliography}
\end{document}